\documentclass{iise}

\usepackage{amsmath,amssymb}
\usepackage{gensymb}

\conference{Proceedings of the IISE Annual Conference \& Expo 2026\\
Y. Xiang, D., Yu, R. Thiesing, eds.}

\title{\titlesize Generative Design for Direct-to-Chip Liquid Cooling for Data Centers}

\author{
Zheng Liu\\University of Michigan-Dearborn\\Dearborn, MI
}
\authorlist{Liu}
\abstractID{17161}

\begin{document}
\maketitle

\begin{abstract}

Rapid growth in artificial intelligence (AI) workloads is driving up data center power densities, increasing the need for advanced thermal management. Direct-to-chip liquid cooling can remove heat efficiently at the source, but many cold-plate channel layouts remain heuristic and are not optimized for the strongly non-uniform temperature distribution of modern heterogeneous packages. This work presents a generative design framework for synthesizing cooling channel geometries for the NVIDIA GB200 Grace Blackwell Superchip. A physics-based finite-difference thermal model provides rapid steady-state temperature predictions and supplies spatial thermal feedback to a constrained reaction–diffusion process that generates novel channel topologies while enforcing inlet/outlet and component constraints. By iterating channel generation and thermal evaluation in a closed loop, the method naturally redistributes cooling capacity toward high-power regions and suppresses hot-spot formation. Compared with a baseline parallel channel design, the resulting channels achieve more than a 5$\degree$C reduction in average temperature and over 35$\degree$C reduction in maximum temperature. Overall, the results demonstrate that coupling generative algorithms with lightweight physics-based modeling can significantly enhance direct-to-chip liquid cooling performance, supporting more sustainable scaling of AI computing.

\end{abstract}

\section*{Keywords}
Generative design, data centers, artificial intelligence, chip cooling, optimization

\section{Introduction}
The rising computational demand from artificial intelligence (AI), cloud services, and high‑performance computing (HPC) is driving rapid data center expansion \cite{dayarathna2015data} and pushing server power densities to levels that challenge traditional air-based thermal management \cite{chang2024liquid}. Cooling remains a major contributor to facility energy use, often exceeding 30\% of total consumption, making thermal control central to both operating cost and sustainability goals \cite{chang2024liquid}. In this context, liquid cooling is gaining momentum as a practical response to the escalating heat fluxes of chips, where conventional air cooling increasingly struggles to maintain safe junction temperatures without resorting to high fan power and impractically high airflow rates \cite{yuksel2021overview}.

Direct-to-chip liquid cooling removes heat from GPU and CPU chips using a coolant with much higher thermal capacity and heat-transfer performance than air \cite{heydari2024experimental}. This enables higher rack densities, tighter temperature control, and potential reductions in overall cooling energy by shifting more of the thermal burden from large-scale airflow management to targeted heat extraction \cite{sharma2012optimal}. It can also improve reliability by reducing thermal cycling and localized temperature spikes, which are common failure accelerators in HPC systems.
However, achieving high-efficiency chip liquid cooling depends strongly on the design of the coolant flow paths \cite{liu2023data,zheng2022electrical}. Prior studies have explored a wide range of cold-plate and microchannel architectures, including straight and serpentine microchannels, tree-like branching networks, V-shaped fin structures, wavy microchannels, double-layer nested microchannel configurations, and hybrid jet-impingement/microchannel designs \cite{wang2025investigation}. Among them, straight channel design have been studied extensively because of their structural simplicity with favorable hydraulic performance. However, in data center hardware, components are often unevenly distributed across the board, such as NVIDIA Blackwell, producing strongly non-uniform heat generation. Straight channel designs, which typically optimized for more uniform heat flux, can therefore suffer reduced cooling effectiveness and localized temperature rise under these spatially varying loads. Thus, better channel design is needed for the data center hardware.

To address these challenges, generative design offers a systematic way to create channel geometries that are tailored to non-uniform chip temperature distribution rather than constrained to a small set of hand-designed layouts \cite{wilson2024generative}. In contrast to traditional parametric tuning (e.g., adjusting channel width, pitch, or manifold dimensions), generative design algorithmically explore a high-dimensional design space to propose channel shapes that balance competing objectives \cite{liu2023generative,sung2024cooling}. These objectives can include minimizing peak temperature and thermal resistance while respecting limits on pressure drop, pumping power, flow maldistribution, and manufacturability constraints. By automatically redistributing coolant toward high-heat regions and shaping flow paths to reduce thermal gradients, generative design can produce non-intuitive channel networks that improve temperature uniformity and cooling efficiency under realistic, spatially varying loads. This study focus on the generative design for the liquid channels for NVIDIA GB200 Grace Blackwell Superchip (based on Grace CPUs and Blackwell GPUs) to reduce the average temperature.

\section{Temperature Simulation}
\begin{figure}[h]
	\centering
	\includegraphics[width=0.9\linewidth]{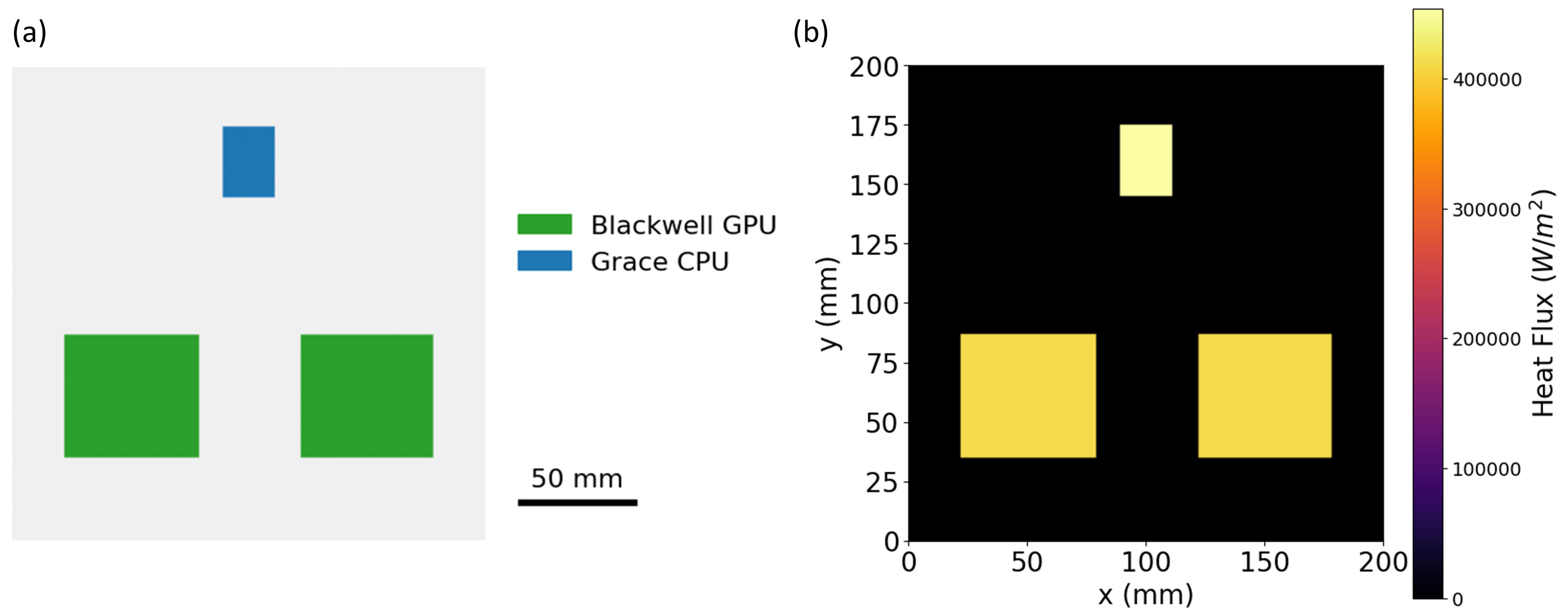}
	\caption{NVIDIA GB200 Grace Blackwell Superchip: (a) Layout, (b) Heat flux.}
    \label{fig:layout}
\end{figure}

In this study, the NVIDIA GB200 Grace Blackwell Superchip is chosen as the target device for the liquid cooling system design. The NVIDIA GB200 Grace Blackwell Superchip consists 2 NVIDIA Blackwell GPUs and 1 NVIDIA Grace CPU \cite{nvidiaNVIDIAGB200}. This study only focus on the area consists chips, as shown in Figure \ref{fig:layout}(a). The thermal design power (TDP) of the GPU is 1,200 W, and the TDP of the CPU is 300 W \cite{fibermallDeepDive}. Because the cold plate is thin and mounted directly to the dies, through-thickness temperature gradients are negligible; therefore, the cooling analysis can be simplified to a 2D steady-state model evaluated under the worst-case TDP heat load. Based on the gemetry and TDP obtained, the heat flux of the design area is shown in Figure \ref{fig:layout}(b).

Because GPUs typically have a tighter allowable temperature rise than CPUs \cite{fibermallDeepDive}, the coolant is introduced on the GPU side and exits on the CPU side to prioritize cooling where thermal margin is most constrained. Both the inlet and outlet are 1 inch (25.4 mm) wide. The simplified simulation can be solved based on Finite Difference Method (FDM), which can solve the 2D steady-state heat conduction equation.

\begin{equation}
k \cdot t \cdot \nabla^2 T + Q(x,y) - h(x,y) \cdot (T - T_{coolant}) = 0
\label{eq:heat}
\end{equation}
where $k$ is the thermal conductivity of silicon ($148~\mathrm{W\,m^{-1}\,K^{-1}}$), $t$ is the chip thickness ($0.001~\mathrm{m}$), $Q(x,y)$ is the heat flux map ($\mathrm{W\,m^{-2}}$), $h(x,y)$ is the local heat transfer coefficient ($\mathrm{W\,m^{-2}\,K^{-1}}$) modulated by the channel mask, $T$ is the temperature, and $T_{\mathrm{coolant}}$ is the inlet coolant temperature ($25^\circ\mathrm{C}$).

The domain can be discretized into a grid with spacing $\Delta x$ and $\Delta y$. And central difference approximations is used \cite{liu2025physics}.

\begin{equation}
\frac{\partial^2 T}{\partial x^2} \approx \frac{T_{i+1,j} - 2T_{i,j} + T_{i-1,j}}{\Delta x^2}, \quad \frac{\partial^2 T}{\partial y^2} \approx \frac{T_{i,j+1} - 2T_{i,j} + T_{i,j-1}}{\Delta y^2}
\label{eq:cd}
\end{equation}

Base on the central difference approximations in Equation~(\ref{eq:cd}), and substitute them in Equation~(\ref{eq:heat}) with defined conduction coefficients $r_x = \frac{k \cdot t}{\Delta x^2}$ and $r_y = \frac{k \cdot t}{\Delta y^2}$, the temperature can be calculated.

\begin{equation}
r_x (T_{i+1,j} + T_{i-1,j}) + r_y (T_{i,j+1} + T_{i,j-1}) - (2r_x + 2r_y + h_{i,j})T_{i,j} + Q_{i,j} + h_{i,j}T_{coolant} = 0
\label{eq:temp}
\end{equation}

The Jacobi method can solve for the temperature at the central node $T_{i,j}$ based on its neighbors.

\begin{equation}
T_{i,j} \cdot (2r_x + 2r_y + h_{i,j}) = r_x(T_{i+1,j} + T_{i-1,j}) + r_y(T_{i,j+1} + T_{i,j-1}) + Q_{i,j} + h_{i,j}T_{coolant}
\label{eq:jacobi}
\end{equation}

Solving for $T_{i,j}$ yields the discrete update rule employed in the numerical implementation.

\begin{equation}
T_{i,j}^{update} = \frac{r_x(T_{left} + T_{right}) + r_y(T_{up} + T_{down}) + Q_{i,j} + h_{i,j}T_{coolant}}{2(r_x + r_y) + h_{i,j}}
\label{eq:update}
\end{equation}

The adiabatic boundary conditions are applied, which assuming no heat flows out of the sides besides the inlet and outlet.

\begin{equation}
\frac{\partial T}{\partial n} = 0
\label{eq:boundary}
\end{equation}

\section{Generative Design}
This work generates candidate cold-plate channel layouts using a constrained reaction-diffusion process. The method leverages the Gray-Scott model \cite{doelman1997pattern} through a thermal feedback loop to produce spatial patterns, while enforcing hard geometric constraints that guarantee channel presence at prescribed inlet/outlet locations and across selected component footprints. Two scalar fields, $U(\mathbf{x},t)$ and $V(\mathbf{x},t)$, evolve according to the Gray-Scott reaction-diffusion system:
\begin{equation}
\label{eq:gs_U}
\frac{\partial U}{\partial t} = D_U \nabla^2 U - U V^2 + F(1-U)
\end{equation}
\begin{equation}
\label{eq:gs_V}
\frac{\partial V}{\partial t} = D_V \nabla^2 V + U V^2 - (F+\kappa)\,V
\end{equation}
where $\Omega \subset \mathbb{R}^2$ denote the design domain, $D_U$ and $D_V$ are diffusion coefficients, $F$ is the feed rate, and $\kappa$ is the kill rate. In the present application, $V$ is treated as a \emph{channel indicator} (larger $V$ corresponding to channel-like regions), while $U$ acts as a complementary field that supports pattern formation through the nonlinear reaction term $UV^2$.

The domain is discretized on a uniform Cartesian grid with spacings $\Delta x$ and $\Delta y$. The Laplacian is approximated using a five-point finite-difference stencil for $\phi \in \{U,V\}$:
\begin{equation}
\label{eq:laplacian_5pt}
\left(\nabla^2 \phi\right)_{i,j} \approx
\frac{\phi_{i+1,j}-2\phi_{i,j}+\phi_{i-1,j}}{(\Delta x)^2}+
\frac{\phi_{i,j+1}-2\phi_{i,j}+\phi_{i,j-1}}{(\Delta y)^2}
\end{equation}

An explicit Euler scheme produces intermediate updates:
\begin{equation}
\label{eq:euler_U}
\tilde U^{n+1}_{i,j} = U^n_{i,j} + \Delta t\left[
D_U\left(\nabla^2 U^n\right)_{i,j} - U^n_{i,j}\left(V^n_{i,j}\right)^2 + F\left(1-U^n_{i,j}\right)
\right]
\end{equation}
\begin{equation}
\label{eq:euler_V}
\tilde V^{n+1}_{i,j} = V^n_{i,j} + \Delta t\left[
D_V\left(\nabla^2 V^n\right)_{i,j} + U^n_{i,j}\left(V^n_{i,j}\right)^2 - (F+\kappa)V^n_{i,j}
\right]
\end{equation}

Homogeneous no-flux boundary conditions are applied on $\partial\Omega$ for the unconstrained evolution and implemented numerically using standard ghost-cell mirroring.
Unconstrained Gray-Scott dynamics can yield disconnected or poorly placed structures. To enforce required geometric features, the restricted method pins $V$ to a prescribed value over selected subsets of grid indices.
The inlet and outlet index sets can be defined:
\begin{equation}
\label{eq:set_in}
\mathcal{P}_{\mathrm{in}} =
\left\{(i,j)\ \middle|\ j=0,\ i_{p0}\le i \le i_{p1}\right\}
\end{equation}
\begin{equation}
\label{eq:set_out}
\mathcal{P}_{\mathrm{out}} =
\left\{(i,j)\ \middle|\ j=N_y-1,\ i_{p0}\le i \le i_{p1}\right\}
\end{equation}

The component-footprint set can be calculated in Equation~(\ref{eq:set_chips}).
\begin{equation}
\label{eq:set_chips}
\mathcal{C}_{\mathrm{chips}} =
\mathcal{C}_{\mathrm{GPU},L}\ \cup\ \mathcal{C}_{\mathrm{GPU},R}\ \cup\ \mathcal{C}_{\mathrm{CPU}}
\end{equation}

Then, the combined pinned set can be defined in Equation~(\ref{eq:set_pinned}).
\begin{equation}
\label{eq:set_pinned}
\mathcal{S}_{\mathrm{pinned}}=
\mathcal{P}_{\mathrm{in}}\cup\mathcal{P}_{\mathrm{out}}\cup\mathcal{C}_{\mathrm{chips}}
\end{equation}

After each reaction-diffusion update, the restriction is enforced as a Dirichlet condition on $V$:
\begin{equation}
\label{eq:pinning_V}
V^{n+1}_{i,j}=
\begin{cases}
1.0, & (i,j)\in \mathcal{S}_{\mathrm{pinned}}\\
\tilde V^{n+1}_{i,j}, & \text{otherwise}
\end{cases}
\end{equation}

Operationally, Equation~(\ref{eq:pinning_V}) acts as a hard geometric constraint that injects persistent of $V$ at the inlet/outlet and across the prescribed component regions, thereby biasing the emergent morphology toward channel networks that connect these required locations.

After convergence, the continuous field $V$ is converted to a binary channel mask $\chi$ using thresholding:
\begin{equation}
\label{eq:threshold_mask}
\chi_{i,j}=
\begin{cases}
1, & V_{i,j}\ge \tau\\
0, & V_{i,j}< \tau
\end{cases}
\end{equation}
where $\tau\in(0,1)$ is a user-defined threshold. The resulting mask can be used to parameterize downstream thermal calculations and to evaluate design objectives such as peak temperature, temperature uniformity, and hydraulic proxies.

\section{Results and Discussion}
\begin{figure}[h]
	\centering
	\includegraphics[width=0.8\linewidth]{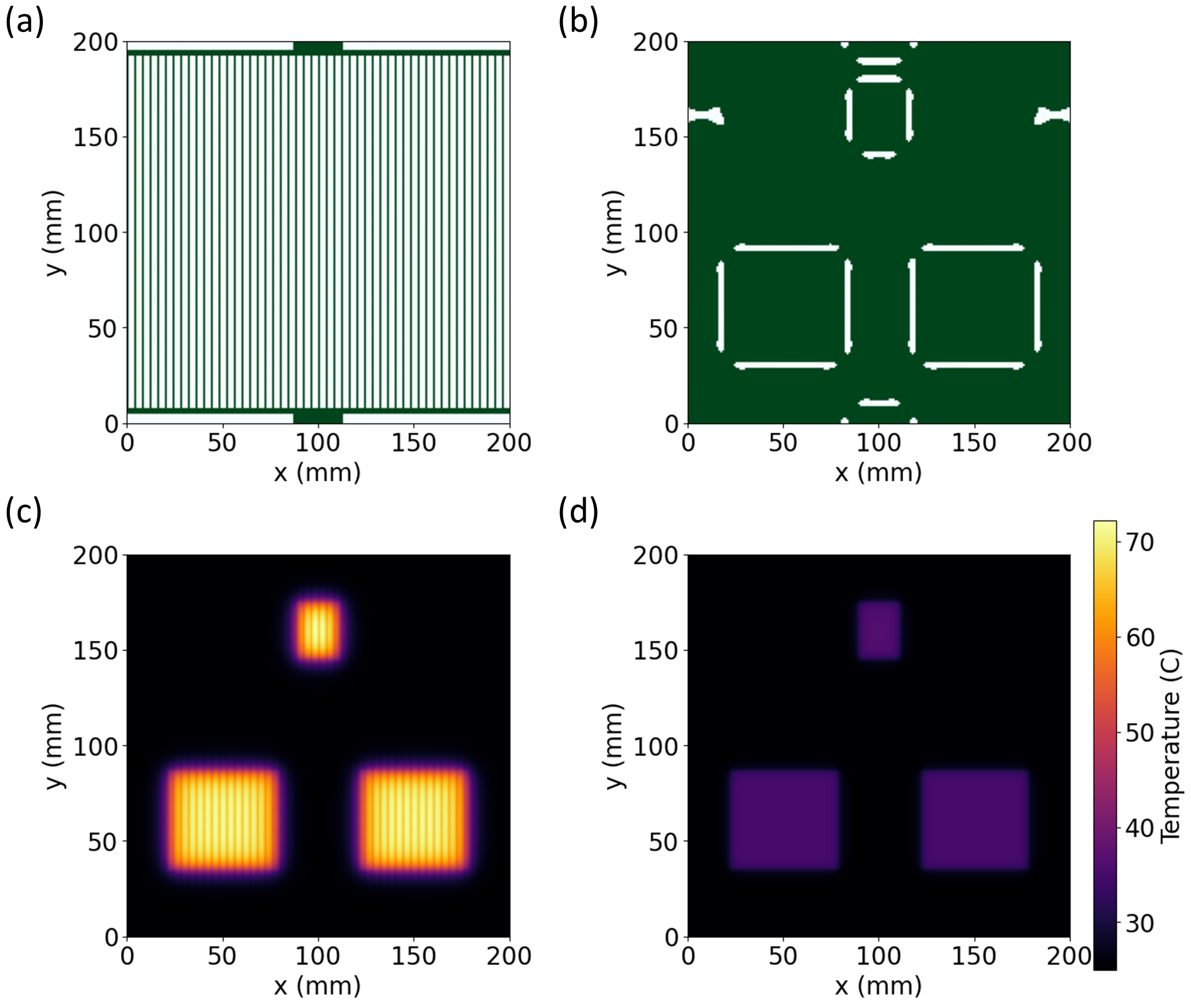}
	\caption{Cooling system layout: (a) Baseline parallel channel design, (b) Generative design. Temperature distribution: (c) Baseline parallel channel design, (d) Generative design.}
    \label{fig:results}
\end{figure}

As shown in Figure \ref{fig:results}, the result of the generative design is compared with baseline of the parallel channels.The comparative data demonstrates a massive performance gap. While the baseline parallel channels struggled to distribute coolant from the narrow ports, resulting in a peak temperature of 72.16$\degree$C, the generative design reduced this to 36.32$\degree$C. While the average temperature of the baseline parallel channels is 32.38$\degree$C, and the average temperature of the generative design reduced this to 26.74$\degree$C.

The baseline’s regular grid layout produced thermal bottlenecks because coolant introduced through the restricted inlet could not redistribute efficiently across the domain. By contrast, the generative approach converged to an organic network that provides efficient branching flow paths from the localized inlet toward the spatially distributed high-power components, thereby alleviating flow maldistribution and reducing temperature non-uniformity. Under stringent geometric constraints, the generative topology approach substantially outperformed conventional Euclidean channel layouts, largely mitigating the thermal penalties that typically arise from restricted coolant delivery.

\bibliographystyle{IEEEtran}
\bibliography{references}

\end{document}